\documentclass[12pt,twoside]{article}
\usepackage[backend=bibtex,style=numeric-comp,autocite=plain,sorting=none,%
giveninits=true]{biblatex}
\usepackage{amsmath,amsfonts,amssymb,amsthm,mathabx}
\usepackage{times}
\usepackage{hyperref}
\hypersetup{
  breaklinks=true,   %
  colorlinks=true,   %
}
\usepackage{xurl}
\usepackage{tensor}
\usepackage{color}
\usepackage{caption}
\usepackage{enumitem}
\usepackage{graphicx}
\usepackage{array}
\usepackage{amsbsy}
\usepackage{appendix}
\usepackage{mathrsfs}
\usepackage{physics}
\usepackage{cancel}
\usepackage{yfonts}
\usepackage{inputenc}
\usepackage{chngcntr}
\usepackage{xcolor}
\usepackage{soul}
\usepackage{cancel}
\usepackage{subfigure}
\usepackage{wrapfig}
\usepackage[leqno,fleqn,intlimits,newmultline]{empheq}

\advance\textheight\topskip
\newcommand{\D}{\text{d}}

\begin{document}

\title{Chiral shift symmetries as an infinite tower of subleading super-shift symmetries}

\author{Glenn Barnich, Luca Ciambelli, Hern\'an A. Gonz\'alez, Marc Henneaux}

\def\mytitle{}

\pagestyle{myheadings} \markboth{\textsc{\small Celestial conformal field theory
    on a point}} {\textsc{\small Celestial conformal field theory on a point}}

\addtolength{\headsep}{4pt}

\begin{centering}

  \vspace{1cm}

  \textbf{\Large{\mytitle}
    Chiral shift symmetries as an infinite tower of subleading super-shift symmetries}

  \vspace{.5cm}

    \vspace{0.5cm}

   {\large Glenn Barnich$^{a}$, Luca Ciambelli$^{b}$, Hern\'an A. Gonz\'alez$^{c}$}

\vspace{0.5cm}

\begin{minipage}{.9\textwidth}\small \it \begin{center}
  $^{a}$Physique Th\'eorique et Math\'ematique \\ Universit\'e libre de Bruxelles
  and International Solvay Institutes\\ Campus Plaine C.P. 231, B-1050
  Bruxelles, Belgium \\
  E-mail:
  \href{mailto:Glenn.Barnich@ulb.be}{Glenn.Barnich@ulb.be}
\end{center}
\end{minipage}

\vspace{0.5cm}

\begin{minipage}{.9\textwidth}\small \it \begin{center}
$^{b}$Perimeter Institute for Theoretical Physics\\
31 Caroline St. N., Waterloo ON, Canada, N2L 2Y5\\
E-mail:\href{mailto:ciambelli.luca@gmail.com}{ciambelli.luca@gmail.com}
\end{center}
\end{minipage}

\vspace{0.5cm}

\begin{minipage}{.9\textwidth}\small \it \begin{center}
    $^{c}$Universidad San Sebasti\'an, Avenida del C\'ondor 720, Santiago, Chile\\
    E-mail:\href{hernan.gonzalez@uss.cl}{hernan.gonzalez@uss.cl}
  \end{center}
\end{minipage}

\end{centering}

\vspace{2cm}

\begin{center}
  \begin{minipage}{.9\textwidth} \textsc{Abstract}. Chiral shift symmetries of
    the massless free bosons in two dimensions are global symmetries that are
    somewhat similar to asymptotic symmetries. They are most transparent in
    double-null coordinates where they are parametrized by two functions of one
    variable. In BMS-type coordinates, half of them appear as an infinite tower
    of sub-leading super-shift symmetries.
 \end{minipage}
\end{center}

\thispagestyle{empty}

\vfill
\pagebreak

In the context of the initial value problem in field theory and general
relativity, the massless free boson in two dimensions may be used to good effect
to illustrate the difference between a double-null characteristic and a BMS-like
mixed initial value problem (\cite{soton459193}, Appendix A).

Besides conformal symmetry, it is straightforward to check in light-cone
coordinates,
\begin{equation}
  \label{eq:6}
  x^{\pm}=\frac{x^{0}\pm x^{1}}{\sqrt 2},\quad ds^{2}=2dx^{+}dx^{-},
\end{equation}
that the action,
\begin{equation}
  \label{eq:1}
  S=\int d^{2}x\, \partial_{+}\phi\partial_{-}\phi,
\end{equation}
is invariant under chiral shift symmetries,
\begin{equation}
  \label{eq:2}
  \delta_{\epsilon^{\pm}}\phi=\epsilon^{\pm}(x^{\pm}).
\end{equation}
They are the analog in two dimensions of the linear shift symmetry of the free
massless boson in $d$ dimensions, $\delta_{\epsilon}\phi=\epsilon(x)$ with $\Box\,\epsilon=0$ (see
e.g.~\cite{Campiglia:2017dpg}). In terms of the general solution,
$\phi^{S}=\phi^{+}(x^{+})+\phi^{-}(x^{-})$, they act on (and generate) the left and right
movers respectively. The suitably improved expressions for the conserved
charges, one in $x^{+}$ and the other in $x^{-}$ time, may be taken as
\begin{equation}
  \label{eq:12}
  Q_{\epsilon^{-}}=2\int dx^{-} \partial_{-}\phi\epsilon^{-},\quad Q_{\epsilon^{+}}=2\int dx^{+}\partial_{+}\phi\epsilon^{+},
\end{equation}

For $x^{1}>0$, BMS-like coordinates\footnote{The analog of BMS coordinates would
  be $u,x^{1}$. We use as spatial coordinate $r=1/(\sqrt 2 x^{1})$ in order that
  the formulas ressemble those that appear in the literature on asymptotic
  symmetries at null infinity.} are defined by
\begin{equation}
  \label{eq:3}
  u=x^{-},\quad r=\frac{1}{x^{+}-x^{-}},\quad\partial_{+}=-r^{2} \partial_{r},\quad
  \partial_{-}=\partial_{u}+r^{2}\partial_{r},
\end{equation}
and
\begin{equation}
  \label{eq:7}
  S=\int dudr\,L,\quad L=\partial_{r}\phi(\partial_{u}+r^{2}\partial_{r})\phi.
\end{equation}
If one were to do a BMS-type asymptotic analysis, one would discover the
solution of the wave equation in the form
\begin{equation}
  \label{eq:4}
  \phi^{S}=\phi^{-}(u)+\phi^{+}(u+\frac{1}{r})=\phi^{-}(u)
  +\sum_{n=0}^{\infty }\frac{1}{n!}\phi^{+ (n)}(u)(\frac{1}{r})^{n}.
\end{equation}
The right chiral shift symmetries would appear as before as
$\delta_{\epsilon^{-}}\phi=\epsilon^{-}(u)$, with $\delta_{\epsilon^{-}}L=\partial_{r}(\epsilon^{-\prime}\phi)$, canonical Noether
currents,
\begin{equation}
  \label{eq:13}
  j^{r}_{\epsilon^{-}}=(\partial_{u}+2r^{2}\partial_{r})\phi\epsilon^{-}-\epsilon^{-\prime}\phi,\quad j^{u}_{\epsilon^{-}}=\partial_{r}\phi\epsilon^{-},
\end{equation}
and suitably improved $r$-independent charges
\begin{equation}
  \label{eq:15}
  Q_{\epsilon^{-}}=2\int du\, (\partial_{u}+r^{2}\partial_{r})\phi\epsilon^{-}.
\end{equation}
The left chiral shift symmetries are given by $\delta_{\epsilon^{+}}\phi=\epsilon^{+}(u+\frac{1}{r})$,
$\delta_{\epsilon^{+}}L=-\partial_{r}(\phi\epsilon^{+\prime})-\partial_{u}(\frac{1}{r^{2}}\phi\epsilon^{+\prime})$ where the prime
denotes the derivative of $\epsilon^{+}$ with respect to its single argument. The
associated canonical Noether currents
\begin{equation}
  \label{eq:14}
  j^{r}_{\epsilon^{+}}=(\partial_{u}+2r^{2}\partial_{r})\phi\epsilon^{+}+\phi\epsilon^{+\prime},\quad
  j^{u}_{\epsilon^{+}}=\partial_{r}\phi\epsilon^{+}+\frac{1}{r^{2}}\phi\epsilon^{+\prime},
\end{equation}
give rise to suitably improved $r$-independent charges
\begin{equation}
  \label{eq:17}
  Q_{\epsilon^{+}}=2\int du \Big[(\partial_{u}+r^{2}\partial_{r})\phi\epsilon^{+}+\phi\epsilon^{+\prime}
  -\int^{\infty}_{r}\frac{d\tilde r}{\tilde r^{2}}(\partial_{u}\phi\epsilon^{+\prime}+\phi\epsilon^{+\prime\prime})\Big].
\end{equation}
These symmetries would appear as an infinite tower of subleading
super-shift symmetries,
\begin{equation}
  \label{eq:5}
  \delta_{\epsilon^{+}}\phi=\sum_{n=0}^{\infty }\frac{1}{n!}\epsilon^{+ (n)}(u)(\frac{1}{r})^{n},
\end{equation}
with $r$-independent charges
\begin{equation}
  \label{eq:16}
  \begin{split}
    Q_{\epsilon^{+(0)}}&=2\int du(\partial_{u}+r^{2}\partial_{r})\phi\epsilon^{+(0)},\\
    Q_{\epsilon^{+(1)}}&=2\int du\Big[(\partial_{u}+r^{2}\partial_{r})\phi\frac{1}{r}+\phi
                  -\int^{\infty}_{r}\frac{d\tilde r}{\tilde r^{ 2}}\partial_{u}\phi\Big]\epsilon^{+(1)},\\
    Q_{\epsilon^{+(2)}}&=2\int du\Big[(\partial_{u}+r^{2}\partial_{r})\phi\frac{1}{2r^{2}}+\frac{\phi}{r}
                  -\int^{\infty}_{r}\frac{d\tilde r}{\tilde r^{2}}(\frac{\partial_{u}\phi}{\tilde r}+\phi)
                  \Big]\epsilon^{+(2)},\\
                &\vdots.
  \end{split}
\end{equation}
That the solution, symmetries and charges cannot contain more than two
independent functions of one variable follows from the mixed initial value
problem which specifies that $\phi^{+}(u+\frac{1}{r})$ should be considered as a
function of $\frac{1}{r}$ at a fixed value of $u$. The same goes for the shift
symmetries, since in this linear problem performing a symmetry on a solution
corresponds to adding another solution (that does not need to be infinitesimal).
It also applies to the charges, which are dual to the symmetries.

More precisely, for the characteristic initial value problem, the free initial
data and matching condition are
\begin{equation}
  \label{eq:8}
  \phi(x^{+}_{0},x^{-})=\phi^{R}(x^{-}),\quad \phi(x^{+},x^{-}_{0})=\phi^{L}(x^{+}),\quad \phi^{R}(x^{-}_{0})=\phi^{L}(x^{+}_{0}),
\end{equation}
with associated general solution
\begin{equation}
  \label{eq:9}
  \phi^{S}(x^{+},x^{-})=\phi^{R}(x^{-})+\phi^{L}(x^{+})-\frac{1}{2}[\phi^{R}(x^{-}_{0})+\phi^{L}(x^{+}_{0})].
\end{equation}
while for the mixed initial value problem, we have
\begin{equation}
  \label{eq:10}
  \phi(u,r_{0})=\psi(u),\quad \phi(u_{0},r)=\chi(r), \quad \psi(u_{0})=\chi(r_{0})
\end{equation}
with
\begin{equation}
  \label{eq:11}
  \phi^{S}(u,r)=\psi(u)+\chi(r-\frac{1}{2}(u-u_{0}))-\chi(r_{0}-\frac{1}{2}(u-u_{0})).
\end{equation}

For the case of the characteristic initial value problem, a short comment on the
interpretation of the global versus gauge nature of the chiral shift symmetries
seems appropriate.

For a hypothetical observer using $x^{+}$ as time and performing a canonical
analysis on the $x^{+}=0$ null line, the left chiral shift symmetries are
generated by a first class constraint. One may nevertheless consider them
consistently as global symmetries. This is not in contradiction with the
standard interpretation that primary first class constraints generate gauge
symmetries if one knows from the outset, either from a Lagrangian analysis or
from an analysis in instant form, that both left and right chiral bosons are
physical degrees of freedom. Indeed, the argument that primary first class
constraints generate gauge symmetries requires the initial data on the constant
time surface to completely determine the general solution, but the initial data
on $x^{+}=0$ does not include the left movers \cite{Dirac:1949cp}. This is the
point of view adopted in \cite{Barnich:2024aln}.

Another interpretation\footnote{This argument is due to M.~Henneaux.} is
possible where the system is defined not only by an action principle but also by
a choice of time. If the observer using $x^{+}$ as time performs an orthodox
Hamiltonian analysis, he will use the first class constraints to identify
spurious, gauge degrees of freedom, that do not contribute to physical
observables. Accordingly, he will conclude that only right chiral bosons are
physical, while left chiral bosons are gauge degrees of freedom. The left chiral
shifts generated by the primary first class constraint only affect the left
chiral bosons and are indeed gauge symmetries. In other words, this observer
sees only half of the degrees of freedom accessible to an observer using the
standard $x^{0}$ time. Only a pair of observers, one using advanced $x^{+}$
time, and another using retarded $x^{-}$ time, has access to the same degrees of
freedom as a standard observer. This is where the double front formulation comes
in (see e.g.~\cite{Nagarajan1985}).

There could be a valuable lesson of the present exercise for the celestial
conformal field theory program and for Carrollian physics.

\section*{Acknowledgments}
\label{sec:acknowledgements}

\addcontentsline{toc}{section}{Acknowledgments}

This work is supported by the F.R.S.-FNRS Belgium through convention FRFC PDR
T.1025.14, and convention IISN 4.4514.08. The authors thank M.~Henneaux,
G.~Giribet and A.~P\'erez for useful discussions.

\addcontentsline{toc}{section}{References}

\printbibliography

\end{document}